\documentclass[twocolumn,english]{aastex631}

\begin{document}
\title{Bottom-up Background Simulations of the 2016 COSI Balloon Flight}

\author[0000-0002-2664-8804]{Savitri Gallego}
\affiliation{Institut für Physik \& Exzellenzcluster PRISMA+,Johannes Gutenberg-Universität Mainz, 55099 Mainz, Germany. email : sgallego@uni-mainz.de}

\author[0000-0001-8160-5498]{Uwe Oberlack}
\affiliation{Institut für Physik \& Exzellenzcluster PRISMA+,Johannes Gutenberg-Universität Mainz, 55099 Mainz, Germany. email : sgallego@uni-mainz.de}

\author[0009-0004-9049-2199]{Jan Lommler}
\affiliation{Institut für Physik \& Exzellenzcluster PRISMA+,Johannes Gutenberg-Universität Mainz, 55099 Mainz, Germany. email : sgallego@uni-mainz.de}

\author[0000-0002-6774-3111]{Christopher M. Karwin}
\affiliation{NASA Goddard Space Flight Center, Greenbelt, MD 20771, USA}

\author[0000-0001-9067-3150]{Andreas Zoglauer}
\affiliation{Space Sciences Laboratory, UC Berkeley, 7 Gauss Way, University of California, Berkeley, CA 94720, USA}

\author[0000-0002-1757-9560]{Pierre Jean}
\affiliation{Institut de Recherche en Astrophysique et Planétologie, Université de Toulouse, CNRS, CNES, UPS, 31028 Toulouse, Cedex 4, France.}

\author[0000-0002-0917-3392]{Peter von Ballmoos}
\affiliation{Institut de Recherche en Astrophysique et Planétologie, Université de Toulouse, CNRS, CNES, UPS, 31028 Toulouse, Cedex 4, France.}

\author[0000-0001-6677-914X]{Carolyn Kierans}
\affiliation{NASA Goddard Space Flight Center, Greenbelt, MD 20771, USA}

\author[0000-0003-4732-6174]{Clio Sleator}
\affiliation{Naval Research Laboratory, 4555 Overlook Ave SW, Washington, DC 20375}

\author[0000-0001-5506-9855]{John A. Tomsick}
\affiliation{Space Sciences Laboratory, UC Berkeley, 7 Gauss Way, University of California, Berkeley, CA 94720, USA}

\author[0000-0001-9567-4224]{and Steven E. Boggs}
\affiliation{Department of Astronomy \& Astrophysics, UC San Diego, 9500 Gilman Drive, La Jolla, CA 92093, USA}

\collaboration{20}{(on behalf of the COSI Collaboration)}



\begin{abstract}

The Compton Spectrometer and Imager (COSI) is a Compton telescope designed to survey the 0.2 - 5 MeV sky, consisting of a compact array of cross-strip germanium detectors. As part of its development, in 2016 COSI had a successful 46 day flight on board NASA’s Super Pressure Balloon platform. This was a precursor to the COSI Small Explorer (COSI-SMEX) satellite mission that will launch in 2027 into an equatorial low Earth (530 km) orbit. The observation of MeV gamma-rays is dominated by background radiation, especially due to the activation of the detector materials induced by cosmic-ray interactions. Thus, background simulation and identification are crucial for the data analysis. Because the COSI-SMEX detectors will be similar to the ones used for the balloon flight, the balloon measurements provide an important tool for testing and cross-checking our background simulations for the upcoming space mission. In this work we perform Monte Carlo simulations of the background emission from the 2016 COSI balloon flight. Including a phenomenological shape correction, we obtain an agreement with the data at the 10-20\% level for energies between 0.1 - 1.6 MeV, and we successfully reproduce most of the activation lines induced by cosmic ray interactions.
\end{abstract}



\section{Introduction} \label{sec:intro}
Balloon campaigns are a key step to developing new experimental
technologies in the gamma-ray domain. Such campaigns have led to numerous fundamental discoveries, including the Crab pulsar~\citep{Fishman1967}, 511~keV emission from the Galactic Center~\citep{511galactic}, nuclear emission lines from the Galactic center region~\citep{Haymes1975DetectionON}, the cosmic gamma-ray background~\citep{osti_5661790}, and the atmospheric MeV background~\citep{EGB1977}. In addition to atmospheric and Cosmic background photons, the background rate of balloon-borne instruments is influenced by the intensity of the incident cosmic-ray radiation that varies with the latitude-dependent cutoff rigidity, altitude and Solar activity. The interactions of these particles not only induce
a prompt instrumental background when they deposit energy in the detectors, but they can also induce a delayed background in soft gamma-ray telescopes due to the activation of the instrument materials. Since gamma-ray measurements are highly dominated by the background in the MeV range, background simulation and identification are crucial for analysis.

In this work we use data from the 2016 balloon flight of the Compton Spectrometer and Imager (COSI) -- a precursor to the COSI Small Explorer (SMEX) satellite mission~\citep{tomsick2023comptonspectrometerimager}. COSI is a Compton telescope which operates as a wide-field imager, spectrometer, and polarimeter. It has an excellent energy resolution (FWHM) of 5.2 keV (7.0 keV) at 0.662 MeV (1.333 MeV)~\citep{Beechert_2022}. Twelve high-purity cross-strip germanium semiconductor detectors (each 8 × 8 × 1.5 cm$^3$) are arranged in a 2 × 2 × 3 array that measures photons between 0.2 and 5 MeV. Six anti-coincidence cesium iodide (CsI) shields surrounding the four sides and bottom of the detector array define the wide $\sim 1 \pi$~sr field of view. The shields suppress the Earth albedo background by actively vetoing particles incident from below the instrument.  

In this work we demonstrate the capability of our background modeling and simulations to reproduce the data energy spectrum and rate, including the delayed background induced by the activation.
The paper is structured as follows. In Section~\ref{sec:COSIflight}, we summarize the 2016 COSI
flight. The background simulations are detailed in Section~\ref{sec:bcksimu}. We illustrate our simulation results with a comparison to the measured data in Section~\ref{sectionresult}. Finally, we discuss our results and give our outlook for the upcoming COSI-SMEX mission in Section~\ref{sec:conclusion}.

\begin{figure}[ht!]
\plotone{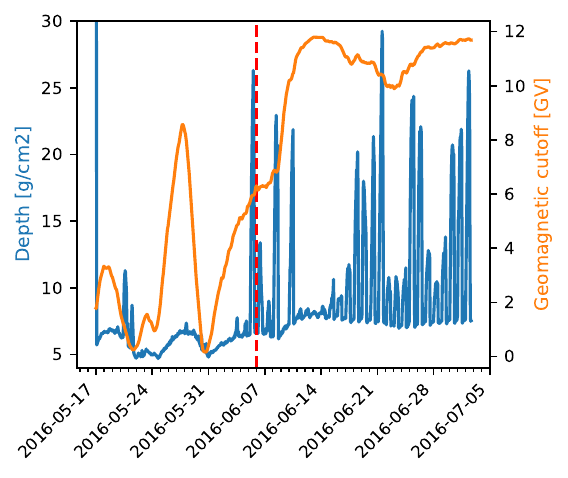}
\caption{Variation of the depth and latitude-dependent cutoff rigidity during the balloon flight. The red dashed line represents the starting date (2016/06/07) of the dataset we present in  Section~\ref{sectionresult}. The spikes in the depth distribution are due to the drop of altitude of the balloon gondola during the nights.}
\label{fig:Balloonparams}
\end{figure}

\section{The 2016 COSI Flight} \label{sec:COSIflight}
On 2016 May 17, COSI was launched as a science payload on a NASA ultra-long duration balloon from Wanaka, New Zealand. The launch site from New Zealand was chosen to maximize exposure of the Galactic center, observations of which are important for COSI’s science goals to measure nuclear lines and electron-positron annihilation. COSI is a freefloating instrument always pointed at zenith and sweeps the sky through the Earth’s rotation during flight. A summary of the 46 day 2016 COSI balloon flight can be found in~\cite{kierans20172016superpressureballoon}. Nine of COSI’s 12 detectors operated continuously throughout the flight. Two detectors were turned off within the first 48 hr of the flight, and a third was turned off on 2016 June 6. The shut-offs were due to a well-understood high voltage problem linked to electronic parts. The nominal flight altitude was 33 km, though the balloon experienced altitude variations between 33 and 22~km with the day-night cycle after 2016 June 4. This corresponds to a depth range from 5 to 30~$\mathrm{g \ cm^{-2}}$. The depth and geomagnetic cutoff variations during the flight are shown in Figure~\ref{fig:Balloonparams}. Remaining at high altitude is preferable for balloon instruments like COSI because the strong atmospheric background decreases with increasing altitude and the cosmic signal is less attenuated. The goal of this work is to account for all of these variations in the background simulations in order to accurately describe the data. 

The instrument circumnavigated the globe within the first 14 days of the flight and then remained largely above the South Pacific Ocean before the flight was safely terminated on 2016 July 2. The instrument was recovered from its landing site in Peru with no signs of consequential damage. For this work we selected data starting from June 7th since the number of working detectors is constant.

\section{Background simulation}\label{sec:bcksimu}

\subsection{Input models}

\begin{figure}[ht!]
\plotone{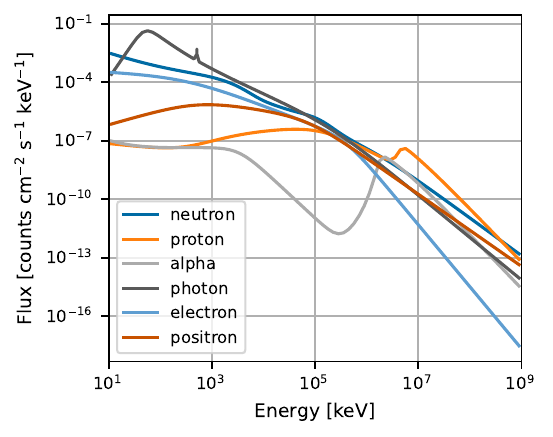}
\caption{EXPACS spectra integrated over all zenith angles. This particular calculation is for the 7th of June 2016, corresponding to a longitude, latitude, depth and cutoff rigidity equal to -100°, -30°, 7~g~cm$^{-2}$ and 5.43~GV, respectively. Each component results from the interactions of cosmic rays in the atmosphere as derived from EXPACS simulations. }
\label{EXPACS_spectrum}
\end{figure}

The intensity and the incident angular distribution of the cosmic rays and secondary particles were   generated using PARMA/EXPACS~\citep{EXPACS}. These fluxes are derived from analytical fits that reproduce extensive air shower simulation results performed with PHITS~\citep{PHITS}. PARMA/EXPACS allows for the generation of fluxes depending on altitude, longitude, latitude, solar activity and geomagnetic cutoff with a zenith angle dependence. In order to account for the variations, we generate particle spectra for every hour of the balloon flight, using the mean altitude, longitude, and latitude during the time bin. An example for June 7th is shown in Figure~\ref{EXPACS_spectrum}.

\begin{figure}[!ht]
\plotone{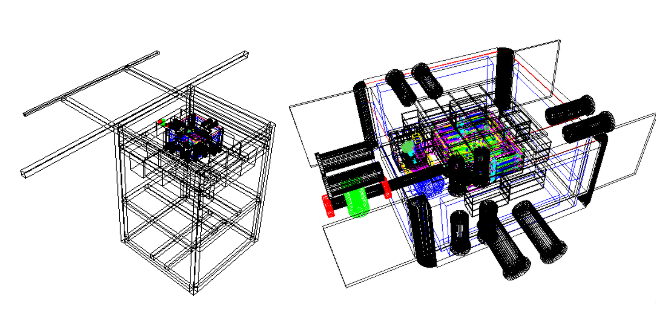}
\plotone{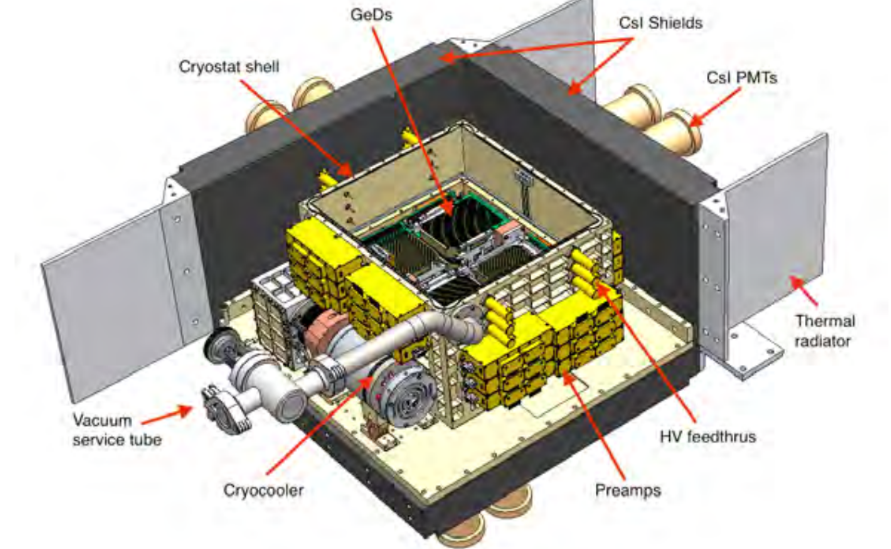}
    \caption{The top left image shows the complete mass model used for the simulations, including the gondola of the balloon and the cryostat. Top right: zoom on the cryostat with Germanium detectors. The bottom image depicts a labeled CAD model of the cryostat in the same orientation with two of the side CsI shields removed. }
    \label{fig:massmodel}
\end{figure}

The simulations employ the Medium-Energy Gamma-ray Astronomy library (MEGAlib) software package, a standard tool in MeV astronomy~\citep{ZOGLAUER2006629}. \textit{Cosima} is the simulation part of MEGAlib, based on GEANT4 v11.1~\citep{GEANT4}. It is able to simulate the emission from a source, the passage of particles through the spacecraft, and their energy deposits in the detector. MEGAlib also performs event
reconstruction, imaging analysis and general high-level analysis (i.e., spectra, light curves, etc.). The mass model used for the simulation is shown in Figure~\ref{fig:massmodel}.

The MEGAlib simulation includes the activation of the instrument induced by cosmic ray interactions. Currently, two options are available for this. The first method is a three step process. First the initial particles and photons are simulated which determines the prompt interactions and a list of the created isotopes are stored in a text file. The activation of each isotope is then calculated based on a given time of irradiation in orbit. And the final step then simulates the isotope decay at random positions within the mass model volumes in which they were created. The second method of activation simulations keeps in memory each isotope created during the simulation until it decays (the expected decay time is computed according to the isotope lifetime and the event is rejected if this time is longer than the simulation time). This latter method accurately simulates the build-up of the activation during the balloon flight, and it has been used for this study.

\begin{figure}[ht!]
\plotone{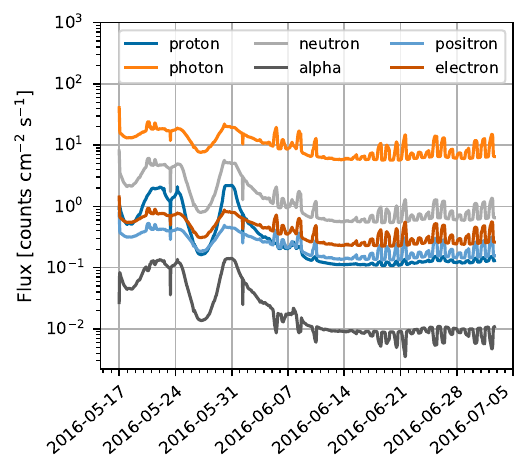}
\caption{Calculated light curves for each background component. The strong variations at the beginning of the flight are due to high variations of the geomagnetic cutoff. The short-duration spikes represent strong variations in altitude}
\label{EXPACS_LC}
\end{figure}

The time-dependence of the background radiation is handled by including a light curve in the simulation. These curves are computed by integrating the model spectra over energy for each hour of the flight and are represented in Figure~\ref{EXPACS_LC}. Representative spectra with zenith angle dependencies are chosen for longitude, latitude, and depth equal to -100°, -30°, and 7~g~cm$^{-2}$, respectively, which is representative of the average values during the second part of the flight (except drops of altitude during the night). For scaling of the flux, we made the simplifying approximation that the spectral shape is constant with time. This is a reasonable assumption for all the components except for the protons and alphas where large geomagnetic cutoff variation in the first part of the flight significantly impacts the spectral shape. However, as we only analyze data after June 7th, only isotopes induced by proton/alpha activation with a lifetime longer than $\sim$5 days could impact the result.

Simulations are run for the full 46 days of the COSI balloon flight in order to take into account the build-up of activation, but the comparison with the data starts only from the 7th of June. The time dependence of the instrument’s pointing on the sky is also simulated.

\subsubsection{Cosmic Diffuse Components}

\begin{figure}[ht!]
\plotone{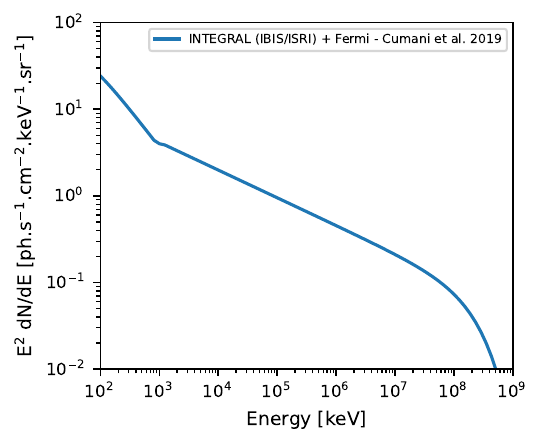}
\caption{Model of the EGB based on fits to data from INTEGRAL-IBIS/ISGRI at low energies, and Fermi-LAT at high energy.}
\label{Cumani_EGB}
\end{figure}

Because PARMA/EXPACS results are derived from simulations of extensive air showers in the atmosphere, the photon component in Figure~\ref{EXPACS_spectrum} does not contain any contribution from the Galactic plane or extragalactic diffuse $\gamma$-ray background. Therefore, they need to be implemented in the background model. Galactic diffuse continuum emission (GDCE) from the COSI ballloon flight was already probed in~\cite{Karwin_2023}. We used the same GALPROP-based model of the Galactic diffuse continuum from that work. For the extragalactic background (EGB) we used the model from~\cite{Cumani_2019}, based on INTEGRAL-IBIS/ISGRI and Fermi-LAT measurements. The resulting spectrum is shown in Figure~\ref{Cumani_EGB}. 

For both the GDCE and the EGB, Earth occultation is accounted for in the simulations by blocking all photons with arrival directions beyond $96^\circ$ of the zenith. We also account for atmospheric scattering. This can be divided into two components. First, a fraction of incident photons will scatter and never reach the detector, causing attenuation of the original signal (the transmitted component). This effect is accounted for in the simulations using a transmission probability file, which has a dependence on the incident photon energy, zenith angle, and detector altitude. The second component consists of photons that reach the detector after scattering one or more times in the atmosphere (the scattered component). This causes an energy-dependent distortion of the measured spectrum, resulting in more photons at lower energy. In order to account for the scattering component we apply the correction factor ratios derived in~\cite{karwin2024atmosphericresponsemevmathrmgamma}. In addition to a dependence on energy and altitude, these correction factors also depend on the zenith angles of the incident photons. As was done in~\cite{karwin2024atmosphericresponsemevmathrmgamma}, we use an off-axis angle of $41^\circ$ for the Galactic diffuse continuum, which is the average angle from the Galactic center during the flight, where the emission is brightest. For the EGB we use an off axis-angle of $45^\circ$. This is the angle where a majority of the photons are detected after accounting for the angular dependence of the effective area combined with the solid angle on the sky, and thus it serves as a reasonable approximation. The correction factor does not drastically change for off-axis angles $<60^\circ$, and thus our assumptions here do not introduce any significant systematic biases. 

\begin{figure}[ht!]
\plotone{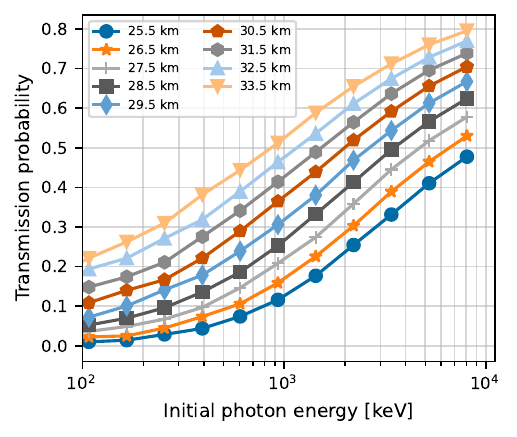}
\plotone{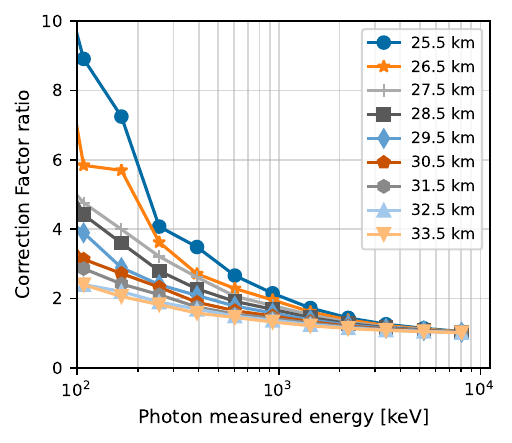}
\caption{\textbf{Top panel:} The atmospheric attenuation for different altitudes corresponding to the transmitted component. \textbf{Bottom panel:} The atmospheric scattering correction factor ratios for different altitudes as defined in~\cite{karwin2024atmosphericresponsemevmathrmgamma} computed for the EGB model.} 
\label{fig:correctionfacratio}
\end{figure}

We include the altitude-dependence of the atmospheric corrections by simulating the balloon flight for altitudes between 25.5 - 33.5 km, with a 1 km spacing, and applying the corrections for each respective altitude (for both the transmitted and scattered components). We then combine the different simulations according to the actual time-dependence of the altitude from the balloon flight. As an example, the correction factor ratios for the EGB at different altitudes are shown in Figure~\ref{fig:correctionfacratio}. The observed counts from the simulations in each respective altitude bin are scaled by this factor in order to account for photons that are scattered into the detector, as described above.

The detector effects engine (DEE) in MEGAlib is applied to the simulated data to account for detector and readout effects. These effects include vetoing events that interact in the shields, dead time, charge sharing, energy calibration, crosstalk correction, strip pairing, etc.~\citep{Clio_paper}. The events are then reconstructed using MEGAlib's \textit{revan}~\citep{revan,ZOGLAUER2006629}. In this study we consider both single-site photoelectric (PE) and Compton (CO) events. For the CO, we apply a selection on data and simulation requiring a minimum distance of 0.5~cm between any interactions.

\section{Results} \label{sectionresult}

\begin{figure}[ht!]
\plotone{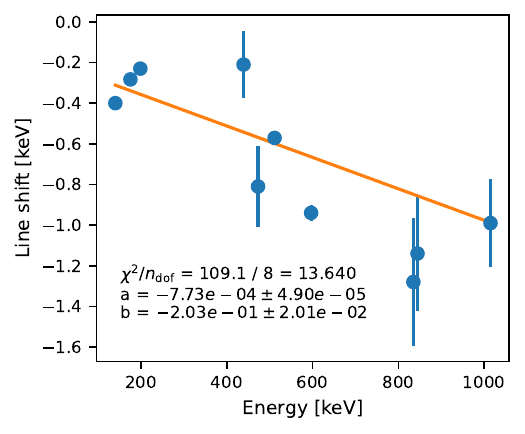}
\caption{Difference of the centroids of the peaks between data (PE) and simulation for some activation lines as function of the energy after applying the additional preamplifier temperature correction. A linear fit $y=a*x+b$ is represented in orange.}
\label{fig:correctionEnergy}
\end{figure}

After passing the simulation files through the DEE and reconstruction pipeline, we observed a systematic shift between the centroid of the activation lines, which increases with energy. It is known that the preamplifier circuits of the Ge detectors are sensitive to temperature and could induce line shifts. This was taken into account and corrected for the balloon flight data \citep{Carolynthesis}. We observed a remnant line shift as a function of preamplifier temperature, although it was very small, around 0.01~keV/°C compared to 0.5~keV/°C before the correction by~\cite{Carolynthesis}. We found an additional correction for the energy shift as a function of preamplifier temperature was needed; however, a line shift as a function of energy is still observed after these temperature corrections. This is represented in Figure~\ref{fig:correctionEnergy}. An additional empirical linear correction was thus applied to each energy deposit in the data in order to better match the line position with the Monte-Carlo simulations. The cause of this remnant shift is still unknown.

After this minor energy correction, the activation lines were better aligned. We then proceeded with our baseline fit in which we scaled the normalizations of the model components to the data using a Poissonian likelihood analysis.  The fit includes normalization factors for the atmospheric photons ($A_{atm}$), hadronic and leptonic components of the internal activation ($A_{int}$), and the cosmic photons, which includes both the Galactic and extragalactic diffuse components ($A_{diff}$).  

\begin{figure}[b!]
\includegraphics{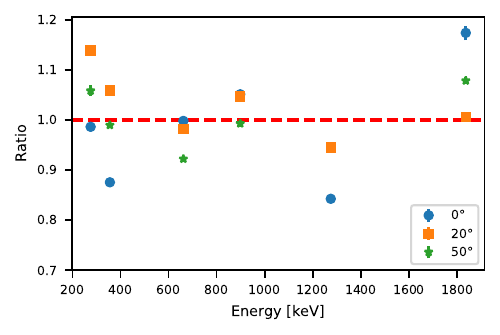}
\includegraphics{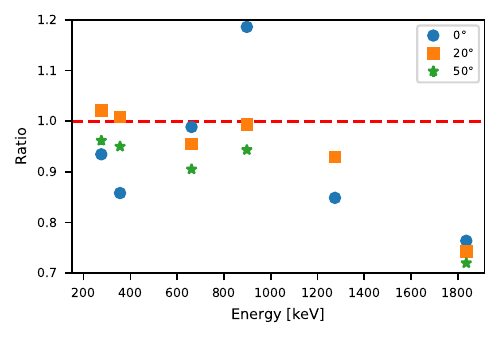}
\caption{A comparison of the ratio of measured to simulated counts in different lines from the calibration campaign as function of energy and off-axis angle. The top panel shows the PE events and the bottom panel shows the CO events. The statistical errors are plotted but they are smaller than the points.}
\label{fig:CalibVSBalloon}
\end{figure}

After performing the baseline fit, we found that the model was systematically overestimating the flux above $\sim$~300-400~keV for both PE and CO events and underestimating the flux below 200~keV for the  CO events. The main effect is seen for the atmospheric photons, since this is the dominant component. However, the slope difference is also observed in other components. This effect on the spectrum is also present when using the Ling model \citep{Ling} for the atmospheric background, instead of the EXPACS model. In addition, as shown in  Figure~\ref{fig:CalibVSBalloon}, the comparison between calibration data and simulation in the photo-peaks indicate a systematic uncertainty of $\sim$10 to 20\% (depending on the energy) from our current DEE. Since the line at 1.8~MeV show the highest uncertainty, we restrict the analysis at 1.6 MeV. However, this is not enough to explain the whole discrepancy with the data. Other effects could intervene such as missing components in the mass model compared to the real instrument.

\begin{figure}[!h]
   \plotone{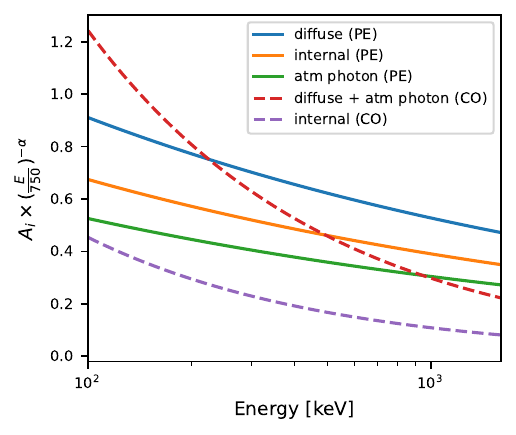}
    \caption{Scaling factors for each components as function of energy. Solid lines are for the PE dataset and dashed-lines are for the CO dataset.}
    \label{fig:scalfac}
\end{figure}

In order to correct for the effects discussed above, we adjusted the slope of the predicted counts by empirically scaling the model components by a power law, defined as $A_i(\frac{E}{E_0})^{-\alpha}$
 with $E_0 = 750$~keV and where $A_i$ is the normalization factor for the respective component and $\alpha$ is a spectral index that is shared for the three components. The components $A_{diff}$ and $A_{atm}$ were merged for the CO dataset because $A_{diff}$ was otherwise tending to 0 in the fit, which is not physical. This is, however, not the case for the PE dataset, where all components are independent and left free. The results of the fits for each dataset are shown in  Table~\ref{tab:fitparameter}, and the corresponding spectral shape compensation are plotted in Figure~\ref{fig:scalfac}.

\begin{table}[!h]
    \begin{tabular}{|c|c|c|}
        \hline
        fit parameter &  PE events & CO events  \\
        \hline
        $A_{atm}$ & 0.326 $\pm$ 0.004   & 0.3562 $\pm$ 0.0016 \\
        $A_{int}$ & 0.4184   $\pm$ 0.0012   &  0.130$\pm$ 0.002 \\
        $A_{diff}$ & 0.565 $\pm$ 0.004 & 0.3562 $\pm$ 0.0016 \\
        $\alpha$ & 0.237 $\pm$ 0.004 & 0.620 $\pm$ 0.001 \\
        \hline
       
    \end{tabular}
    
    \caption{Best-fit parameters for both PE and CO events.}
     \label{tab:fitparameter}
\end{table} 

\begin{figure*}[p]
\centering
\includegraphics[trim=15 15 10 10,clip,width=0.75\textwidth]{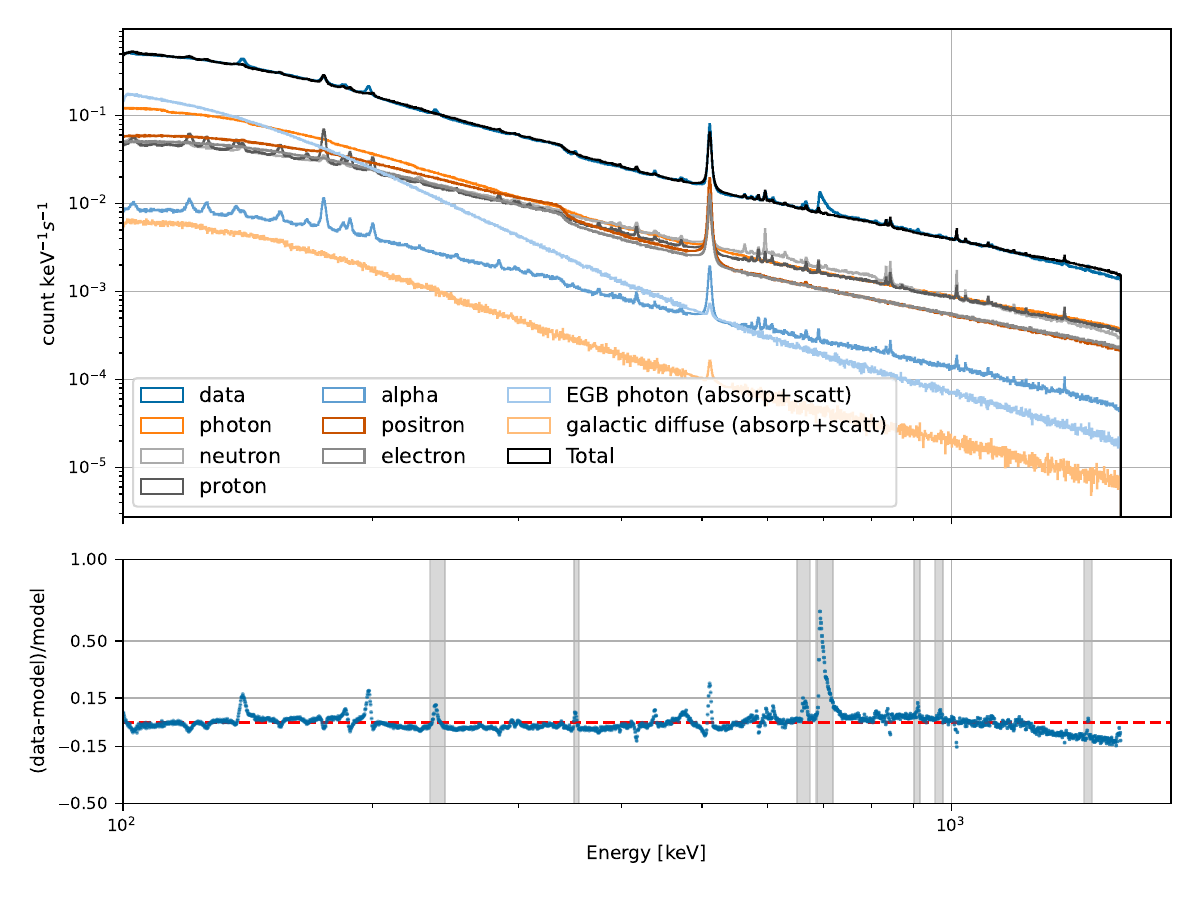}
\caption{\textbf{Top:} Spectra of the PE events for all components of the model. \textbf{Bottom:} Residuals between data and total background model. The grey areas show activation lines present in the data but not in the simulations (or vice-versa).}
\label{fig:PEeventSpec}

\includegraphics[trim=15 15 10 10,clip,width=0.75\textwidth]{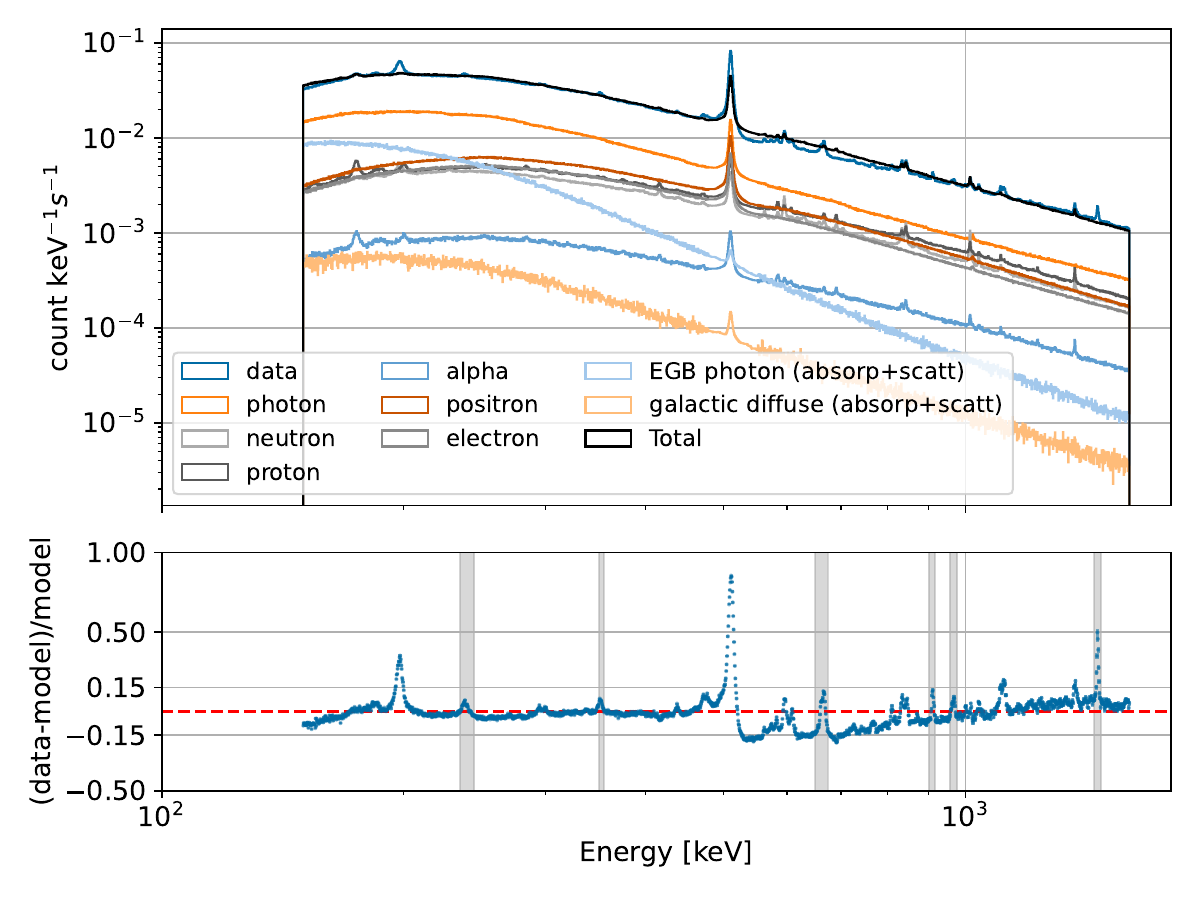}
\caption{\textbf{Top:} Spectra of the Compton events for all components of the model. \textbf{Bottom:} Residuals between data and total background model. The grey areas show activation lines present in the data but not in the simulations (or vice-versa).}
\label{fig:COeventSpec}
\end{figure*}

The comparisons between the total background model and the data  for both the PE and CO events are shown in Figures~\ref{fig:PEeventSpec} and~\ref{fig:COeventSpec}. Each component includes all interactions within the germanium detector induced by the incident particles. This includes prompt interactions such as bremsstrahlung, as well as interactions resulting from activation. After empirical correction, we obtained a 10 to 20\% agreement with the measurement. The compton edge of the 511~keV line at $\sim$340~keV, clearly visible in the data for the PE dataset is nicely reproduce by the simulations. 
Most of the activation lines are quite well simulated. Some of them, however, are observed in the data but not in our background model. They are written in the top panel of Table~\ref{tab:absentlines} with their corresponding parent process. 

It appears that the lines that are observed in the data but not in the simulations originate from natural radioactivity, mostly from the $^{232}$Th and $^{238}$U series. We checked that these lines are present in the background measurement taken during the calibration campaign. An example of this measurement compared to the balloon data is shown in Figure~\ref{fig:datavsbckmeas}. Because of the natural decay origin of those lines, their absence in our simulation is expected. The line present in the PE dataset around 694~keV with a triangular shape is known to be due to the reaction $^{72}$Ge(n,n')$^{72}$Ge$^*$ \citep{BOGGS2002390}. This peculiar shape derives from the fact that this reaction is completely based on internal transition. This is also the reason why we only observe this line in the PE events dataset but not the Compton one. The reason why we do not observe this line in our neutron component seems to be due to a very specific recent GEANT4 issue where for some cases of internal transition, a gamma-ray is emitted instead of an electron \citep{geant4issue}.  Since this line is only present in the PE events dataset, modeling of the CO events is unaffected.

\begin{figure*}[!h]
\includegraphics[trim=0 5 0 5,clip]{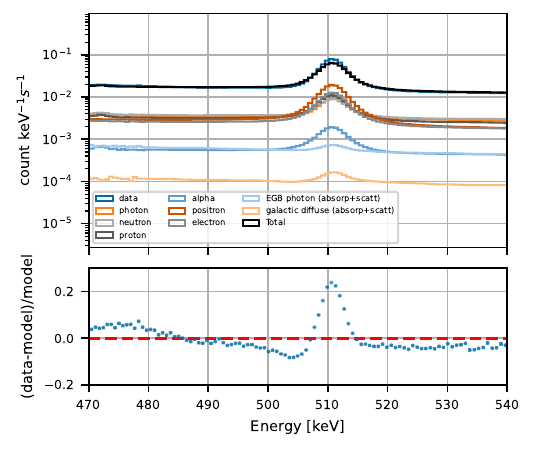}
\includegraphics[trim=0 5 0 5,clip]{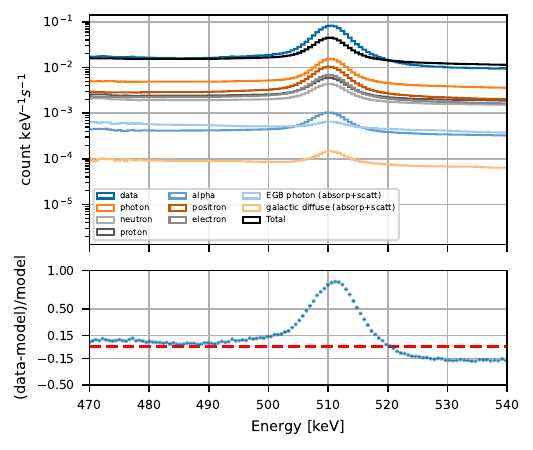}
\caption{\textbf{Left: } Zoom of Figure~\ref{fig:PEeventSpec} for the PE events around the 511~keV line. \textbf{Right: } Similar zoom for the CO events.  }
\label{fig:PEeventSpec_511zoom}
\end{figure*}

\begin{figure}[h]
\includegraphics[trim=0 5 0 5,clip]{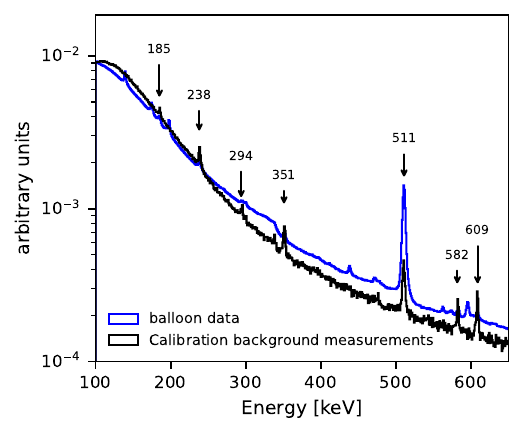}
\caption{Spectrum of the PE events for the balloon data from 2016/06/07 to 2016/07/01 in blue. The natural radioactivity background measurement during the calibration campaign is shown in black. Spectra are normalized to their area. Arrows indicate lines present in both spectra.}
\label{fig:datavsbckmeas}
\end{figure}

\begin{figure*}[h]
\centering
\includegraphics[trim=20 8 20 75, clip, width=0.9\textwidth]{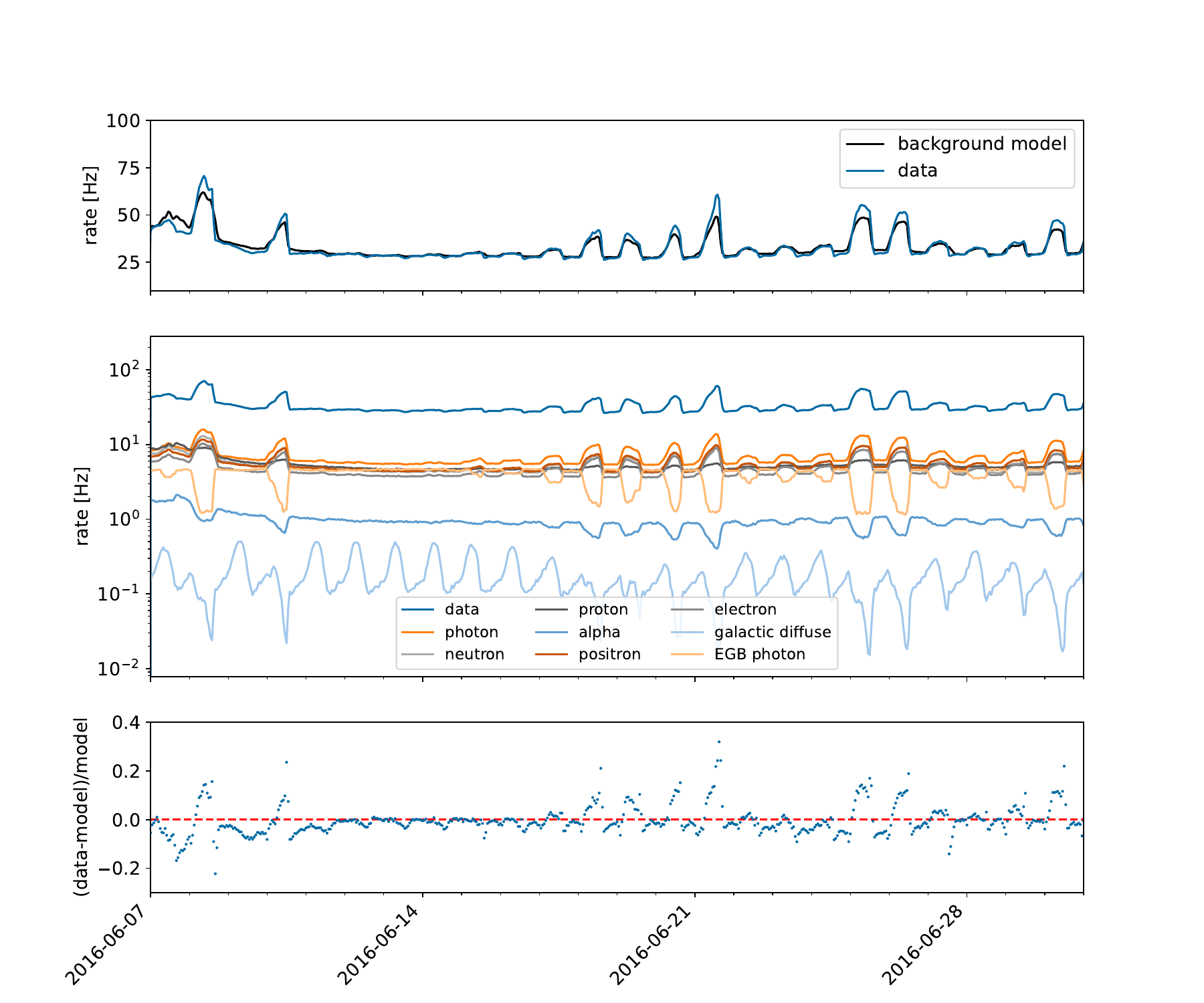}
\caption{\textbf{Top:} Rate comparison of the PE events between data (blue) and the total background model (in black) for the period 2016/06/07 to 2016/07/01. The rate is integrated between 0.1 and 1.6~MeV. \textbf{Middle: } Similar to the upper plot but for each component of the model. \textbf{Bottom:} Residuals between data and total background model.}
\label{fig:PETotalRate}
\end{figure*}

For all the lines we observed a greater width in the data than in the simulation, resulting in a higher intensity in the simulation. This effect, clearly visible in the residuals, has been seen and quantified during the benchmarking of the DEE \citep{Clio_paper}.   
This broadening effect is even higher for the 511~keV line. One of the reasons for the hadronic/leptonic background components is the fact that in GEANT4 the electron-positron pair that produces the 511~keV $\gamma$-rays is at rest, when in reality the electron and positron have some momentum; this non-zero momentum leads to the broadening of the line in the data, also called Doppler broadening. This enhanced effect at 511~keV has been seen during the calibration when comparing simulation with measurement for a $^{22}$Na source (see Fig 16 in \cite{Clio_paper}). For the atmospheric photon component, Compton scattering of the 511~keV gamma-ray in the atmosphere could lead to a broadening of the line that would change with altitude~\citep{Harris2003}. This effect is not taken into account in the simulations. Finally, the broadening of the 511~keV line in the data could also come at some level from Galactic positron annihilation~\citep{Prantzos,Siegert23}. However, this last contribution should be negligible giving the fact we are not making any pointing or selection cuts. In order to correct the simulation for the broadening of the 511~keV line, we fit this line in the data with a convolution of a normal and a Cauchy distribution, also called a voigt profile. The fit result give a sigma of 2.39 for the normal distribution and 2.03 for the FWHM of the Cauchy distribution for both PE and CO dataset. This type of shape works well for representing the Doppler broadening of positron annihilation~\citep{511shape}. The Voigt profile was then convolved with the 511~keV line from the simulation. This approach allows us to smooth the line, similar to applying a Gaussian filter. The results are shown in Figure~\ref{fig:PEeventSpec_511zoom}.
The integrated rate for the PE dataset is shown in Figure~\ref{fig:PETotalRate}. It is notable that the "standard" rate intervals are fit very well, whereas for the periods of low altitude(high rates) simulated rates fall a bit short of the data. This could be explain by known limitations with incomplete volume descriptions in the balloon instrument mass model.

\renewcommand\floatpagefraction{1}
\renewcommand\topfraction{1}

\begin{table*}[h!]
    \centering
    \begin{tabular}{|c|c|c|c|}
       \hline
       Energy measured [keV]  & Nominal energy [keV]& Parent T$_{1:2}$ & Parent process \\
       \hline
        238.35(3) & 238.632(2) & 10.622 h &$^{212}Pb(\beta^-)^{212}Bi \hspace{0.1cm}(^{232}Th \hspace{0.1cm}\text{series})$\\
        295.15(1) & 295.2228(18) & 27.06 min &$^{214}Pb(\beta^-)^{212}Bi \hspace{0.1cm}(^{238}U  \hspace{0.1cm}\text{series})$\\
        351.4(6) & 351.9321(18) & 27.06 min  &$^{214}Pb(\beta^-)^{212}Bi \hspace{0.1cm}(^{238}U  \hspace{0.1cm}\text{series})$\\
        661.74(19) & 661.7 & 30.08 y   &$^{137}Cs(\beta^-)^{137}Ba$\\
        910.62(24) & 911.204(2) & 6.15 h & $^{228}Ac(\beta^-)^{228}Th \hspace{0.1cm}(^{238}U  \hspace{0.1cm}\text{series})$\\
        961.7(10) & 960.67(5) & 1.71 h  & $^{202}Bi(EC)^{202}Pb$\\
        967(1) & 964.777(11) or 968.974(17) &  6.15 h & $^{228}Ac(\beta^-)^{228}Th \hspace{0.1cm}(^{238}U  \hspace{0.1cm}\text{series})$\\
        1460.76(12) & 1460.820(5) & 1.248e9 y & $^{40}K(EC)^{40}Ar$\\
        \textcolor{blue}{693.94(26)} & \textcolor{blue}{691.3} &  ?& $^{72}Ge(n,n')^{72}Ge$\\

   \hline\hline  
   139.4(4) & 139.68(3)& 47.7 s & $^{75m}Ge(IT)^{75}Ge$\\
   174.73(20)& 174.949(4)& 81 ns& $^{71m}Ge(IT)^{71}Ge$\\
   184.94(35)& 184.577(10)&3.2617 d &$^{67}Ga(IT)^{67}Zn$ \\
   198.00(11)& 198.392(16)& 20.22 ms&$^{71m}Ge(IT)^{71}Ge$ \\
   438.01(14)& 438.634(18)&13.756 h &$^{69m}Zn(IT)^{69}Zn$ \\
   471.91(21)& 472.202(9)&20.18 ms &$^{24m}Na(IT)^{24}Na$ \\
   562.78(11)&562.93(3) &18.14 ps & $^{76}Ge(n,n')$\\
   573.33(15)&574.17(3) &39.05 h &$^{69}Ge(EC)^{69}Ga$ \\
   582.86(32)&584.54(3) &39.05 h &$^{69}Ge(EC)^{69}Ga +K$ \\
   595.75(10)&595.847(6) &12.41 ps & $^{74}Ge(n,\gamma)^{74}Ge$\\
   667.16(20)&666.94(6) & ? &$^{64m}Ga(EC)^{64}Ga$ \\
   833.9(1)&834.848(3) & 312.20 d& $^{54}Mn(EC)^{54}Cr$\\
   844.02(14)&843.74(3) &9.435 min &$^{27}Mg(\beta^-)^{27}Al$ \\
   1014.44(26)&1014.42(3) & 9.435 min & $^{27}Mg(\beta^-)^{27}Al$\\
   1039.7(34)&1039.513(10) &9.304 h & $^{66}Ga(EC)^{66}Zn$\\
   1107.06(14)&1108.2 &39.05 h & $^{69}Ge(EC)^{69}Ga + L$ \\
   1116.43(20)& 1117.38(6)&39.05 h & $^{69}Ge(EC)^{69}Ga + K$\\
   1369.48(26)& 1368.626(5)& 14.96 h&$^{24}Na(\beta^-)^{24}Mg$ \\
   \hline
    \end{tabular}
    \caption{\textbf{Top: }Table of activation lines that are present in the data but not the simulation due to their natural radioactivity origin. The line in blue is problematic due to some GEANT4 issues during internal transition as described in~\citep{geant4issue}. \textbf{Bottom: } Similar table but for the other activation lines that are both present in data and simulation. }
    \label{tab:absentlines}
\end{table*}

\section{Summary, Discussion, and Conclusion} \label{sec:conclusion}

In this work, we simulated the complete 2016 COSI balloon flight, taking into account the variation of altitude and attitude. Our background model reaches a satisfactory agreement with the measurements at the 10-20\% level for energies between 0.1-1.6 MeV, and we successfully reproduce most of the activation lines induced by cosmic ray interactions. The broadening of the 511~keV line still needs to be better understood, especially the impact of the atmosphere response and the contribution from Galactic positron annihilation. Variations of the total rate with altitude are also reproduced at the 20\% level. However, this work also points to the need for a better understanding of the response of our detector in order to extend our model to 5~MeV.

Our modeling of the COSI background is useful for developing the background model for COSI-SMEX. We have already performed the background simulations for COSI-SMEX, using 3 months of exposure, with an equatorial orbit at an altitude of 550~km\footnote{The target altitude changed from 550 km to 530 km to avoid the Starlink shells.}, and a zenith pointing. This was performed as part of the second \href{https://doi.org/10.5281/zenodo.15126188}{Data Challenge} released by the COSI collaboration. This also coincided with the first alpha release of the high-level analysis software being developed, cosipy~\citep{COSIpy}. In general, the purpose of these data challenges is to aid in the pipeline development, and to train the astrophysics community on how to analyze COSI data. Although the backgrounds at low-Earth orbit differ from the atmospheric backgrounds at balloon altitudes, the activation lines and background components will be similar.

The work presented in this paper provides the prospect that the background simulation of COSI-SMEX will be able to predicts the future measurements close to the level of uncertainties inherent in the input models, thanks to the progress in GEANT4 and MEGAlib and to the detailed studies possible with the balloon flight and calibration data. It is, however, also clear that substantial work still lies ahead, as a new detector effect engine and detailed mass models are being developed, and tuned with calibrations.

\section*{Acknowledgments}
The Compton Spectrometer and Imager is a NASA Explorer project led by
the University of California, Berkeley with funding from NASA under contract
80GSFC21C0059. The COSI-balloon program was supported through NASA-APRA grants NNX14AC81G and 80NSSC19K1389.

C.M.K.’s research was supported by an appointment to the NASA Postdoctoral Program at NASA Goddard Space Flight Center, administered by Oak Ridge Associated Universities under contract with NASA.

SG and JL acknowledge support by DLR grant 50OO2218. Resources supporting this work were provided by National High Performance Computing (NHR) South-West at Johannes Gutenberg University Mainz.

This work is also supported in part by the Centre National d’Etudes Spatiales (CNES)

%

\vspace{5mm}
\facilities{COSI}







\begin{thebibliography}{}
\expandafter\ifx\csname natexlab\endcsname\relax\def\natexlab#1{#1}\fi
\providecommand{\url}[1]{\href{#1}{#1}}
\providecommand{\dodoi}[1]{doi:~\href{http://doi.org/#1}{\nolinkurl{#1}}}
\providecommand{\doeprint}[1]{\href{http://ascl.net/#1}{\nolinkurl{http://ascl.net/#1}}}
\providecommand{\doarXiv}[1]{\href{https://arxiv.org/abs/#1}{\nolinkurl{https://arxiv.org/abs/#1}}}

\bibitem[{Agostinelli {et~al.}(2003)Agostinelli, Allison, Amako, Apostolakis, Araujo, Arce, Asai, Axen, Banerjee, Barrand, Behner, Bellagamba, Boudreau, Broglia, Brunengo, Burkhardt, Chauvie, Chuma, Chytracek, Cooperman, Cosmo, Degtyarenko, Dell'Acqua, Depaola, Dietrich, Enami, Feliciello, Ferguson, Fesefeldt, Folger, Foppiano, Forti, Garelli, Giani, Giannitrapani, Gibin, {Gómez Cadenas}, González, {Gracia Abril}, Greeniaus, Greiner, Grichine, Grossheim, Guatelli, Gumplinger, Hamatsu, Hashimoto, Hasui, Heikkinen, Howard, Ivanchenko, Johnson, Jones, Kallenbach, Kanaya, Kawabata, Kawabata, Kawaguti, Kelner, Kent, Kimura, Kodama, Kokoulin, Kossov, Kurashige, Lamanna, Lampén, Lara, Lefebure, Lei, Liendl, Lockman, Longo, Magni, Maire, Medernach, Minamimoto, {Mora de Freitas}, Morita, Murakami, Nagamatu, Nartallo, Nieminen, Nishimura, Ohtsubo, Okamura, O'Neale, Oohata, Paech, Perl, Pfeiffer, Pia, Ranjard, Rybin, Sadilov, {Di Salvo}, Santin, Sasaki, Savvas, Sawada, Scherer, Sei, Sirotenko, Smith, Starkov,
  Stoecker, Sulkimo, Takahata, Tanaka, Tcherniaev, {Safai Tehrani}, Tropeano, Truscott, Uno, Urban, Urban, Verderi, Walkden, Wander, Weber, Wellisch, Wenaus, Williams, Wright, Yamada, Yoshida, \& Zschiesche}]{GEANT4}
Agostinelli, S., Allison, J., Amako, K., {et~al.} 2003, Nuclear Instruments and Methods in Physics Research Section A: Accelerators, Spectrometers, Detectors and Associated Equipment, 506, 250, \dodoi{https://doi.org/10.1016/S0168-9002(03)01368-8}

\bibitem[{Beechert {et~al.}(2022)Beechert, Lazar, Boggs, Brandt, Chang, Chu, Gulick, Kierans, Lowell, Pellegrini, Roberts, Siegert, Sleator, Tomsick, \& Zoglauer}]{Beechert_2022}
Beechert, J., Lazar, H., Boggs, S.~E., {et~al.} 2022, Nuclear Instruments and Methods in Physics Research Section A: Accelerators, Spectrometers, Detectors and Associated Equipment, 1031, 166510, \dodoi{10.1016/j.nima.2022.166510}

\bibitem[{Boggs {et~al.}(2002)Boggs, Jean, Slassi-Sennou, Coburn, Lin, Madden, McBride, Pelling, Primbsch, \& {von Ballmoos}}]{BOGGS2002390}
Boggs, S., Jean, P., Slassi-Sennou, S., {et~al.} 2002, Nuclear Instruments and Methods in Physics Research Section A: Accelerators, Spectrometers, Detectors and Associated Equipment, 491, 390, \dodoi{https://doi.org/10.1016/S0168-9002(02)01228-7}

\bibitem[{{Boggs, S. E.} \& {Jean, P.}(2000)}]{revan}
{Boggs, S. E.}, \& {Jean, P.} 2000, Astron. Astrophys. Suppl. Ser., 145, 311, \dodoi{10.1051/aas:2000107}

\bibitem[{Cumani {et~al.}(2019)Cumani, Hernanz, Kiener, Tatischeff, \& Zoglauer}]{Cumani_2019}
Cumani, P., Hernanz, M., Kiener, J., Tatischeff, V., \& Zoglauer, A. 2019, Experimental Astronomy, 47, 273–302, \dodoi{10.1007/s10686-019-09624-0}

\bibitem[{Delgado(2023)}]{geant4issue}
Delgado, A.~T. 2023, {Problem 2566 - Converted electrons are emitted as gammas}, \url{https://bugzilla-geant4.kek.jp/show_bug.cgi?id=2566}

\bibitem[{{do Nascimento} {et~al.}(2005){do Nascimento}, Helene, Takiya, \& Vanin}]{511shape}
{do Nascimento}, E., Helene, O., Takiya, C., \& Vanin, V. 2005, Nuclear Instruments and Methods in Physics Research Section A: Accelerators, Spectrometers, Detectors and Associated Equipment, 538, 723, \dodoi{https://doi.org/10.1016/j.nima.2004.09.013}

\bibitem[{Gerald {et~al.}(1969)Gerald, Harnden, Johnson, \& Haymes}]{Fishman1967}
Gerald, F., Harnden, J., Johnson, I., \& Haymes. 1969, The Astrophysical Journal, 158, L61, \dodoi{10.1086/180432}

\bibitem[{Harris {et~al.}(2003)Harris, Share, \& Leising}]{Harris2003}
Harris, M.~J., Share, G.~H., \& Leising, M.~D. 2003, Journal of Geophysical Research: Space Physics, 108, \dodoi{https://doi.org/10.1029/2003JA009958}

\bibitem[{Haymes {et~al.}(1975)Haymes, Walraven, Meegan, Hall, Djuth, \& Shelton}]{Haymes1975DetectionON}
Haymes, R.~C., Walraven, G.~D., Meegan, C.~A., {et~al.} 1975, The Astrophysical Journal, 201, 593.
\newblock \url{https://api.semanticscholar.org/CorpusID:122030389}

\bibitem[{{Johnson} \& {Haymes}(1973)}]{511galactic}
{Johnson}, III, W.~N., \& {Haymes}, R.~C. 1973, \apj, 184, 103, \dodoi{10.1086/152309}

\bibitem[{Karwin {et~al.}(2024)Karwin, Kierans, Shih, Castellanos, Lowell, Siegert, Roberts, Gallego, Laviron, Zoglauer, Tomsick, \& Boggs}]{karwin2024atmosphericresponsemevmathrmgamma}
Karwin, C., Kierans, C., Shih, A., {et~al.} 2024, Atmospheric Response for MeV $\mathrm{\gamma}$ Rays Observed with Balloon-Borne Detectors.
\newblock \doarXiv{2406.03534}

\bibitem[{Karwin {et~al.}(2023)Karwin, Siegert, Beechert, Tomsick, Porter, Negro, Kierans, Ajello, Martinez-Castellanos, Shih, Zoglauer, Boggs, \& (for~the COSI~Collaboration)}]{Karwin_2023}
Karwin, C.~M., Siegert, T., Beechert, J., {et~al.} 2023, The Astrophysical Journal, 959, 90, \dodoi{10.3847/1538-4357/ad04df}

\bibitem[{Kierans(2018)}]{Carolynthesis}
Kierans, C. 2018, PhD thesis, UC Berkeley.
\newblock \url{https://escholarship.org/uc/item/1244t3h7}

\bibitem[{Kierans {et~al.}(2017)Kierans, Boggs, Chiu, Lowell, Sleator, Tomsick, Zoglauer, Amman, Chang, Tseng, Yang, Lin, Jean, \& von Ballmoos}]{kierans20172016superpressureballoon}
Kierans, C.~A., Boggs, S.~E., Chiu, J.-L., {et~al.} 2017, The 2016 Super Pressure Balloon flight of the Compton Spectrometer and Imager.
\newblock \doarXiv{1701.05558}

\bibitem[{Ling(1975)}]{Ling}
Ling, J.~C. 1975, Journal of Geophysical Research (1896-1977), 80, 3241, \dodoi{https://doi.org/10.1029/JA080i022p03241}

\bibitem[{Marshall {et~al.}(1980)Marshall, Boldt, Holt, Miller, Mushotzky, Rose, Rothschild, \& Serlemitsos}]{osti_5661790}
Marshall, F.~E., Boldt, E.~A., Holt, S.~S., {et~al.} 1980, Astrophys. J.; (United States), 235:1, \dodoi{10.1086/157601}

\bibitem[{Martinez(2023)}]{COSIpy}
Martinez, I. 2023, in Proceedings of 38th International Cosmic Ray Conference — PoS(ICRC2023), ICRC2023 (Sissa Medialab), \dodoi{10.22323/1.444.0858}

\bibitem[{Prantzos {et~al.}(2011)Prantzos, Boehm, Bykov, Diehl, Ferri\`ere, Guessoum, Jean, Knoedlseder, Marcowith, Moskalenko, Strong, \& Weidenspointner}]{Prantzos}
Prantzos, N., Boehm, C., Bykov, A.~M., {et~al.} 2011, Rev. Mod. Phys., 83, 1001, \dodoi{10.1103/RevModPhys.83.1001}

\bibitem[{Sato {et~al.}(2008)Sato, Yasuda, Niita, Endo, \& Sihver}]{PHITS}
Sato, T., Yasuda, H., Niita, K., Endo, A., \& Sihver, L. 2008, Radiation Research, 170, 244 , \dodoi{10.1667/RR1094.1}

\bibitem[{Schönfelder {et~al.}(1977)Schönfelder, Graser, \& Daugherty}]{EGB1977}
Schönfelder, V., Graser, U., \& Daugherty, J. 1977, The Astrophysical Journal, 217, 306, \dodoi{10.1086/155580}

\bibitem[{Siegert(2023)}]{Siegert23}
Siegert, T. 2023, Astrophys Space Sci 368, 27 (2023)., \dodoi{10.1007/s10509-023-04184-4}

\bibitem[{Sleator {et~al.}(2019)Sleator, Zoglauer, Lowell, Kierans, Pellegrini, Beechert, Boggs, Brandt, Lazar, Roberts, Siegert, \& Tomsick}]{Clio_paper}
Sleator, C.~C., Zoglauer, A., Lowell, A.~W., {et~al.} 2019, Nuclear Instruments and Methods in Physics Research Section A: Accelerators, Spectrometers, Detectors and Associated Equipment, 946, 162643, \dodoi{https://doi.org/10.1016/j.nima.2019.162643}

\bibitem[{Tomsick {et~al.}(2023)Tomsick, Boggs, Zoglauer, Hartmann, Ajello, Burns, Fryer, Karwin, Kierans, Lowell, Malzac, Roberts, Saint-Hilaire, Shih, Siegert, Sleator, Takahashi, Tavecchio, Wulf, Beechert, Gulick, Joens, Lazar, Neights, Oliveros, Matsumoto, Melia, Yoneda, Amman, Bal, von Ballmoos, Bates, Böttcher, Bulgarelli, Cavazzuti, Chang, Chen, Chu, Ciabattoni, Costamante, Dreyer, Fioretti, Fenu, Gallego, Ghirlanda, Grove, Huang, Jean, Khatiya, Knödlseder, Krause, Leising, Lewis, Lommler, Marcotulli, Martinez-Castellanos, Mittal, Negro, Nussirat, Nakazawa, Oberlack, Palmore, Panebianco, Parmiggiani, Parsotan, Pike, Rogers, Schutte, Sheng, Smale, Smith, Trigg, Venters, Watanabe, \& Zhang}]{tomsick2023comptonspectrometerimager}
Tomsick, J.~A., Boggs, S.~E., Zoglauer, A., {et~al.} 2023, The Compton Spectrometer and Imager.
\newblock \doarXiv{2308.12362}

\bibitem[{T.Sato(2016)}]{EXPACS}
T.Sato. 2016, PLOS ONE, 11(8), e0160390.
\newblock \url{https://doi.org/10.1371/journal.pone.0160390}

\bibitem[{Zoglauer {et~al.}(2006)Zoglauer, Andritschke, \& Schopper}]{ZOGLAUER2006629}
Zoglauer, A., Andritschke, R., \& Schopper, F. 2006, New Astronomy Reviews, 50, 629, \dodoi{https://doi.org/10.1016/j.newar.2006.06.049}

\end{thebibliography}

\bibliographystyle{aasjournal}



\end{document}